\documentclass[11pt,aps,preprint,tightenlines,showpacs,groupadress]{revtex4}
\usepackage{graphicx}
\usepackage{rotating}
\usepackage{amssymb}
\usepackage{mathptmx}
\usepackage{tabularx}

\makeatletter
\bibpunct{}{}{,}{s}{}{,}

\begin{document}

\newcommand{\hdblarrow}{H\makebox[0.9ex][l]{$\downdownarrows$}-}
\title{Condensate fraction in liquid $^{\bf 4}$He at zero temperature}

\author{R. Rota$^1$ and J. Boronat$^1$}

\affiliation{1:Departament de F\'\i sica i Enginyeria Nuclear, \\
Campus Nord B4-B5, Universitat Polit\`ecnica de Catalunya, \\
08034 Barcelona, Spain\\
%Tel.:\\ Fax:\\
\email{jordi.boronat@upc.edu}}

\date{28.09.2011}

\begin{abstract}
We present results of the one-body density matrix $\rho_1(r)$
and the condensate fraction $n_0$ of liquid $^4$He calculated at zero
temperature by means of the Path Integral Ground State Monte Carlo
method. This technique allows to generate a highly accurate
approximation for the ground state wave function $\Psi_0$ in a
totally model-independent way, that depends only on
the Hamiltonian of the system and on the symmetry properties of
$\Psi_0$. With this unbiased estimation of $\rho_1(r)$, we obtain precise results for the
condensate fraction $n_0$ and the kinetic energy $K$ of the
system. The dependence of $n_0$ with the pressure shows an excellent agreement of our results with recent
experimental measurements. Above the melting pressure, overpressurized liquid $^4$He shows a small condensate fraction that has dropped to $0.8\%$ at the highest pressure of $p = 87 \, {\rm bar}$.\\
{\bf keywords}: Liquid Helium, Bose-Einstein condensation, Quantum Monte Carlo.\\
\end{abstract}


\maketitle

\section{Introduction}
Several microscopic theories point out that the phenomenon of superfluidity
in liquid $^4$He has to be seen as a consequence of Bose-Einstein
condensation (BEC).\cite{TilleyTilley} Having total spin $S = 0$, $^4$He
atoms behave like bosons and, below the critical temperature $T_{\lambda} =
2.17 \, {\rm K}$, they can occupy macroscopically the same single-particle
state. Nevertheless, the strong interaction between $^4$He atoms does not
allow all of them to occupy the lowest energy state and, even at zero
temperature, only a small fraction $n_0 = N_0/N$ of the $N$ particles is in
the condensate.

The macroscopic occupation of the lowest energy state, in a strongly
correlated system like $^4$He, appears in the momentum distribution
$n(\bf{k})$ as a delta-peak at $\bf{k} = 0$ and a divergent behavior $n(k)
\sim 1/k$ when $k \to 0$. In the coordinate space, the presence of BEC in a
homogeneous system can be deduced from the asymptotic behavior of the
one-body density matrix $\rho_1(\bf{r})$ ($n_0 =
\lim_{r\to\infty}\rho_1(r)$), which is the inverse Fourier transform of
$n(\bf{k})$.

Experimental estimates of $n_0$ can be obtained from the dynamic structure
factor, $S(q,\omega)$, measured by neutron inelastic scattering at high
energy and momentum transfer. These measurements have a long history:\cite{GlydeBook,SilverSokol,SvenssonSeara} in the 80s, the first
experiments gave estimates for $n_0$ slightly above 10\%, but they were
affected by a poor instrumental resolution and by some difficulties in
describing the final states effects of the scattering experiment. Recently,
with the advances in the experimental technology and in the method of
analysis of the scattering data, Glyde {\it et al.}\cite{Glyde00} have been
able to give improved estimations of $n_0$ at very low temperature. At
saturated vapor pressure (SVP), they found $n_0 = (7.25 \pm 0.75)
\%$,\cite{Glyde00} and more recently they have measured the dependence of
$n_0$ with pressure $p$.\cite{Glyde11}

Because of the strong correlations between $^4$He atoms, the calculation of
the one-body density matrix in superfluid $^4$He cannot be obtained
analytically via a perturbative approach. It is necessary the use of
microscopic simulations to provide accurate estimations of the condensate
fraction. In particular, the Path Integral Monte Carlo (PIMC) method has
been widely used in the study of $^4$He at finite temperature, thanks to
its capability of furnishing in principle exact numerical estimates of
physical observables relying only on the Hamiltonian of the
system.\cite{CeperleyRevPIMC} The first calculations of $n_0$ with this
method date back to 1987,\cite{Ceperley87} but most recent simulations
based on an improved sampling algorithm provide very accurate results for
$\rho_1(r)$, showing a condensate fraction $n_0 = 0.081 \pm 0.002$ at
temperature $T = 1 \, {\rm K}$. \cite{BoninsegniWorm} At zero temperature,
ground-state projection techniques are widely used in the study of BEC
properties of $^4$He. Diffusion Monte Carlo technique, for instance, has
provided estimations of $n_0$ in liquid $^4$He on a wide range of
pressures.\cite{Boronat94,Moroni97,Vranjes05} This method, however, suffers
from the choice of a variational ansatz necessary for the importance
sampling whose influence on $\rho_1(r)$ cannot be completely removed.
Reptation Quantum Monte Carlo (RQMC) has also been used for this
purpose,\cite{Moroni04} but the calculated value of $n_0$ at SVP lies
somewhat below the recent PIMC  value \cite{BoninsegniWorm} at $T = 1 \,
{\rm K}$ noted above.

Motivated by recent accurate experimental data on $n_0(p)$, our aim in the
present work is to perform new calculations of $\rho_1(r)$ and of $n_0$ in
liquid $^4$He at zero temperature using a completely model-independent
technique based on path integral formalism. The Path Integral Ground State
(PIGS) Monte Carlo method is able to compute exact quantum averages of
physical observables without importance sampling, that is without taking
into account any {\it a priori} trial wave function.\cite{Rossi09} Using a
good sampling scheme in our Monte Carlo simulations, we are able to provide
very precise calculations of the one-body density matrix at different
densities. We fit our numerical data for $\rho_1(r)$ with the model used in
previous experimental works, \cite{Glyde00} highlighting the merits and the
faults of this model, and finally we give our estimations for the
condensate fraction when changing the pressure of the liquid, showing an
excellent agreement with experimental data.\cite{Glyde11}

The PIGS method and the computational details of our simulation are
discussed in Sec. \ref{Sec_PIGS}. The results are presented in Sec.
\ref{Sec_Results} and Sec. \ref{Sec_Conclusions} comprises the main
conclusions.

\section{The PIGS method}\label{Sec_PIGS}
The PIGS approach to the study of quantum systems consists in a systematic
improvement of a trial wave function $\Psi_T$ by repeated application of the
evolution operator in imaginary time, which eventually drives the system
into the ground state,\cite{SarsaPIGS} according to the formula
\begin{equation}\label{Eq_PIGSwf}
 \Psi_{PIGS}(R_M) = \int \prod_{i=1}^{M} dR_{i-1} G(R_i,R_{i-1};\tau) \Psi_T(R_0) \ ,
\end{equation}
where the $R_i = \{ {\bf r}_{i;1},{\bf r}_{i;2},...,{\bf r}_{i;N} \}$ represent different sets of coordinates the $N$ particles of the system, and $G(R',R;\tau)=\langle R' \vert e^{-\tau\hat{H}}\vert R \rangle$ is the imaginary time propagator.

Given an approximation of $G(R,R';\tau)$ for small $\tau$, the averages of diagonal
observables can be calculated mapping the quantum many-body system onto a
classical system made up of $N$ interacting polymers composed by $2 M +1$ beads, each of them representing a different evolution in imaginary time of the initial trial state $\Psi_T$. Increasing the number $M$, one is able to
reduce the systematic error and therefore to recover ''exactly'' the ground-state properties of the system.

A good approximation for the propagator $G$ is
important for improving the numerical efficiency of the method:
this greatly reduces the complexity of the calculation
and ergodicity issues, allowing to simulate the quantum system with few
beads, each one with a large time step. Using a high-order approximation
for the propagator, it is possible to obtain an accurate description of the
exact ground state wave function with little numeric effort, even when the
initial trial wave function contains no more information than the bosonic
statistics, that is when one starts the
imaginary time evolution from $\Psi_T=1$.\cite{Rota10}

The one-body density matrix can be written as
\begin{equation}\label{Eq_OBDMPsi}
\rho_1({\bf r}_1,{\bf r'}_1) = \frac{\int d{\bf r}_2 ... d{\bf r}_N
\Psi^*_0(R)\Psi_0(R')}{\int d{\bf r}_1 ... d{\bf r}_N
|\Psi_0(R)|^2} \ ,
\end{equation}
where the configuration $R = \{ {\bf r}_1,{\bf r}_2,...,{\bf r}_N
\}$ differs from $R' = \{ {\bf r'}_1,{\bf r}_2,...,{\bf r}_N \}$
only by the position of one of the $N$ atoms. In the PIGS approach, the
expectation value of non-diagonal observables, like $\rho_1$, is computed
mapping the quantum system in the same classical system of polymers as in
the diagonal case, but cutting one of these polymers in the mid point.
Building the histogram of the frequencies of the distances between the cut
extremities of the two half polymers, one can compute the numerator in Eq.
(\ref{Eq_OBDMPsi}). The calculation of the normalization factor at the
denominator is not strictly necessary since the histogram can be normalized
imposing the condition $\rho_1(0)=1$. However, this {\it a posteriori}
normalization procedure is not easy, because of the small occurrences of
the distances close to zero, and may introduce systematic errors in the
estimation of the condensate fraction $n_0$. In our work, we have avoided
this problem incorporating in the sampling the worm algorithm (WA), a
technique previously developed for path integral Monte Carlo simulations at
finite temperature.\cite{BoninsegniWorm} The key aspect of WA is to work in
an extended configuration space, containing both diagonal (all polymers
with the same length) and off-diagonal (one polymer cut in two separate
halves) configurations, and one of its main advantages is its capability of
evaluating the normalization factor in off-diagonal estimators. We have
extended this technique to zero-temperature calculations and we have been
able to get automatically the properly normalized $\rho_1$, and therefore
very precise estimations of $n_0$. In our simulations, the sampling also
contains movements involving the principle of indistinguishability of
quantum particles, like the {\it swap} update \cite{BoninsegniWorm}. Even
though swaps are not strictly necessary since boson symmetry is fulfilled
with a proper choice of the imaginary time propagator, they improve the
sampling allowing a larger displacement of the half polymers and thus a
better exploration of the long range limit of $\rho_1$.

\section{Results}\label{Sec_Results}
\begin{figure}
\includegraphics[width=\linewidth]{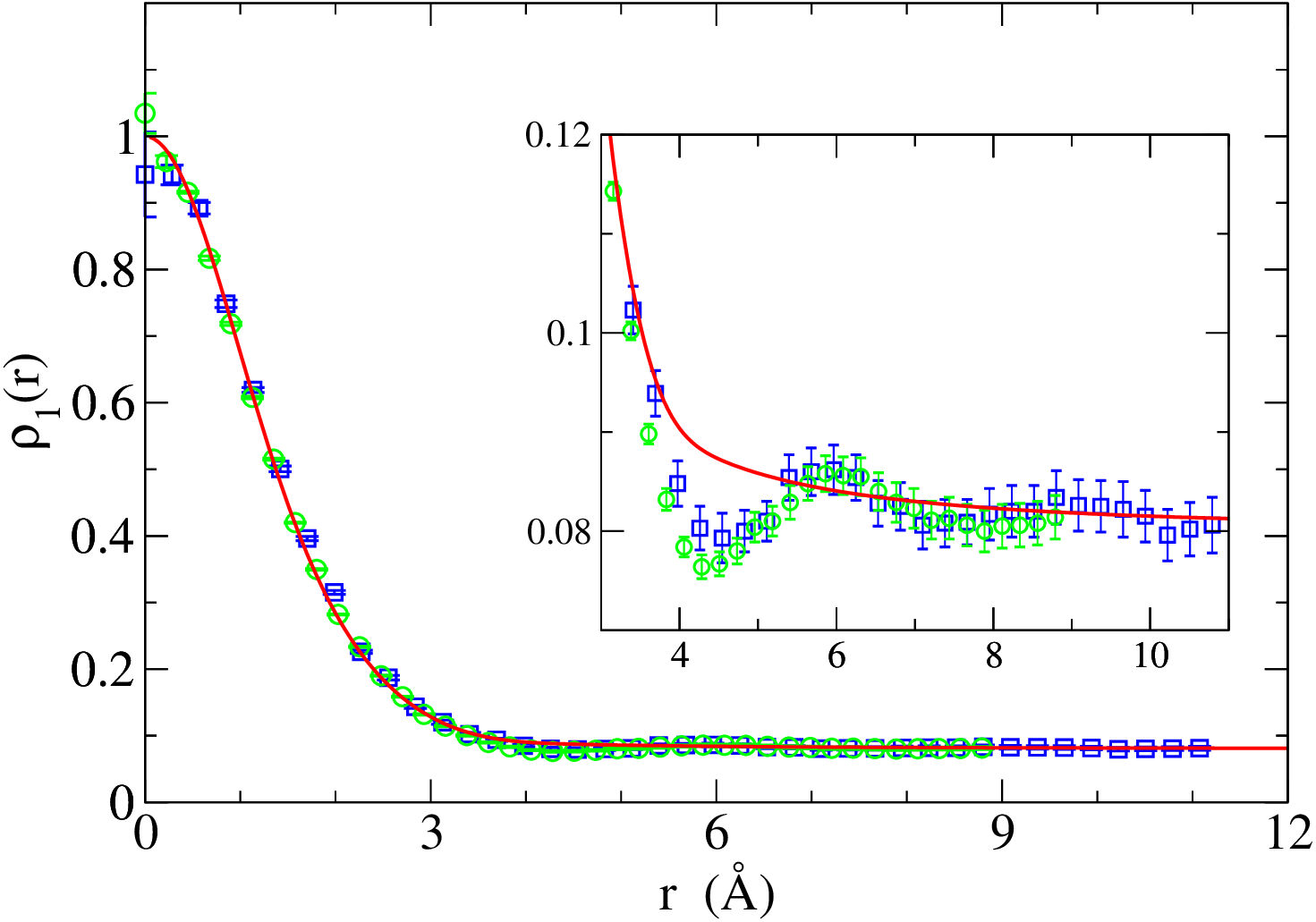}
\caption{(Color online) One-body density matrix $\rho_1(r)$ at the
equilibrium density $\rho = 0.02186$ \AA$^{-3}$. The symbols
represents the result of the PIGS simulations for the system containing
$N = 128$ (green circles) and $N = 256$ (blue squares) $^4$He atoms. The
red line is the curve obtained fitting these data with Eq.
\ref{Eq_OBDM_fit} with optimal values: $k_c=1.369 \pm 0.020\ \textrm{\AA}^{-1}$, $\alpha_2=0.794
\pm 0.005\ \textrm{\AA}^{-2}$, $ \alpha_4=0.355 \pm 0.050\ \textrm{\AA}^{-4}$, $\alpha_6=0.680 \pm 0.080\ \textrm{\AA}^{-6}$, and
$n_0=0.0801 \pm 0.0022$. The inset shows the same data for $r$ between 3
\AA and
11 \AA on an expanded scale.}\label{FigOBDM_eq}
\end{figure}

To compute $\rho_1(r) = \rho_1(|{\bf r}_1 - {\bf r'}_1|)$ in liquid $^4$He
at several densities, we have carried out different simulations with a
cubic box with periodic boundary conditions containing $N = 128$ atoms
interacting through the Aziz pair potential.\cite{Aziz} At first we study
the system at the equilibrium density $\rho = 0.02186 \,$ \AA$^{-3}$:
our result for $\rho_1(r)$ is shown in Fig. \ref{FigOBDM_eq}. We have
checked that our results starting with $\Psi_T = 1$ or with a
Jastrow-McMillan wave function are statistically indistinguishable. In
order to check how the finite size of the box affects our results, we have
performed a simulation of the same system in a larger box containing $N=256$
$^4$He atoms. In Fig. \ref{FigOBDM_eq}, we have compared $\rho_1(r)$
obtained in this last simulation with the one estimated using a smaller
number of particles: we can see that, up to the distances reachable with
the smaller system, these two results agree within the statistical error.
Furthermore, the two functions reach the same plateau at
the largest available distances, indicating that
the asymptotic regime for $\rho_1(r)$ is already achieved using $N=128$
$^4$He atoms.

To fit our data we use the model proposed by Glyde in Ref.
\cite{GlydeBook} that has been used in the analysis of
experimental data,\cite{Glyde00}
\begin{equation}\label{Eq_OBDM_fit}
\rho_1(r) = n_0[1+f(r)]+ A \rho_1^*(r) \ .
\end{equation}
The function $f(r)$ represents the coupling between the condensate and
the non-zero momentum states. In momentum space, one can express $f(k)$ in
terms of the phonon response function \cite{GlydeBook},
\begin{equation}\label{Eq_fk}
f(k) = \left[\frac{m c}{2 \hbar (2 \pi)^3 \rho} \frac{1}{k} \coth \left(
\frac{c \hbar k}{2 k_B T} \right)\right] e^{-k^2/(2 k_c^2)} \ ,
\end{equation}
with $c$ the speed of sound.
Since we work in the coordinate space, we are interested in its 3D
Fourier transform $f(r)$, which at zero temperature can be
written as
\begin{equation} f(r) = \frac{m c}{\hbar (2 \pi)^2 \rho}
\frac{\sqrt{2} k_c}{r} D\left( \frac{k_c r}{\sqrt{2}} \right) \ ,
\end{equation}
where $D(x) = e^{-x^2} \int_{0}^{x}{dt e^{t^2}}$ is the Dawson function.
To describe the contribution to
$\rho_1$ from the states above the condensate, which we denote by
$\rho_1^*$, we use the cumulant expansion of the intermediate
scattering function, that is the Fourier Transform of the
longitudinal momentum distribution,\cite{GlydeBook}
\begin{equation}\label{Eq_OBDM_noncond} \rho_1^*(r) = \exp
\left[ -\frac{\alpha_2 r^2}{2!}+ \frac{\alpha_4 r^4}{4!}
-\frac{\alpha_6 r^6}{6!}\right] \ .
\end{equation}
The constant $A$ appearing in Eq. (\ref{Eq_OBDM_fit}) is fixed by the
normalization condition $\rho_1(0) = 1$. Therefore,
the model we used has five parameters: $n_0$, $k_c$,
$\alpha_2$, $\alpha_4$ and $\alpha_6$. It has to be noticed that,
unlike what is done in the treatment of the experimental data,
where $k_c$ is chosen as a cut-off parameter to make the term $f(k)$
vanish out of the phonon region, we have considered $k_c$ as a free parameter
of the fit.

The best fit we get using the model of Eq. (\ref{Eq_OBDM_fit}) is shown
in Fig. \ref{FigOBDM_eq}. This model is able to reproduce the behavior of
$\rho_1(r)$ for short distances
and in the asymptotic regime, but cannot describe well the
numerical data in the range of intermediate $r$. Indeed, for
distances above 3 \AA, $\rho_1(r)$ obtained with PIGS presents
oscillations which are damped at larger $r$, as observed also in
previous theoretical calculations.\cite{Moroni04,BoninsegniWorm}
This non monotonic behavior, which can be attributed to coordination shell 
oscillations,\cite{BoninsegniWorm} is difficult to describe within the
model of Eq. \ref{Eq_OBDM_fit}.

Nevertheless, despite of these difficulties in describing the
oscillations of the $\rho_1(r)$ obtained with PIGS, the fit we
gave using the model of Eq. (\ref{Eq_OBDM_fit}) contains important
informations about the ground state of liquid $^4$He. First of
all, from the long range behavior, we can obtain the value of the
condensate fraction $n_0$. From our analysis, we get $n_0 = 0.0801
\pm 0.0022$, in complete agreement with the value $n_0 = 0.081 \pm
0.002$ obtained by Boninsegni {\it et al.}\cite{BoninsegniWorm} in
a path integral Monte Carlo simulation at temperature $T = 1 \,
{\rm K}$, and in good agreement with the experimental result $n_0
= 0.0725 \pm 0.0075$.\cite{Glyde00} Furthermore, from the behavior
of $\rho_1(r)$ at short distances, we can obtain an estimation of
the kinetic energy per particle $K/N$. In particular, the term
$\alpha_2$ appearing in Eq. (\ref{Eq_OBDM_noncond}) is the second
moment of the struck atom wave vector projected along the
direction of the incoming neutron \cite{GlydeBook} and is related
to the kinetic energy per particle by the formula $K/N = 3
(\hbar^2/2m) \alpha_2$. Using the value $\alpha_2 = (0.794 \pm
0.005)$ \AA$^{-2}$ obtained in our fit, we get $K/N = (14.43 \pm
0.09) \, {\rm K}$ which has to be compared with the value obtained
in the PIGS simulation, $K/N = (14.37 \pm 0.03) \, {\rm K}$.

\begin{figure}
\includegraphics[width=\linewidth]{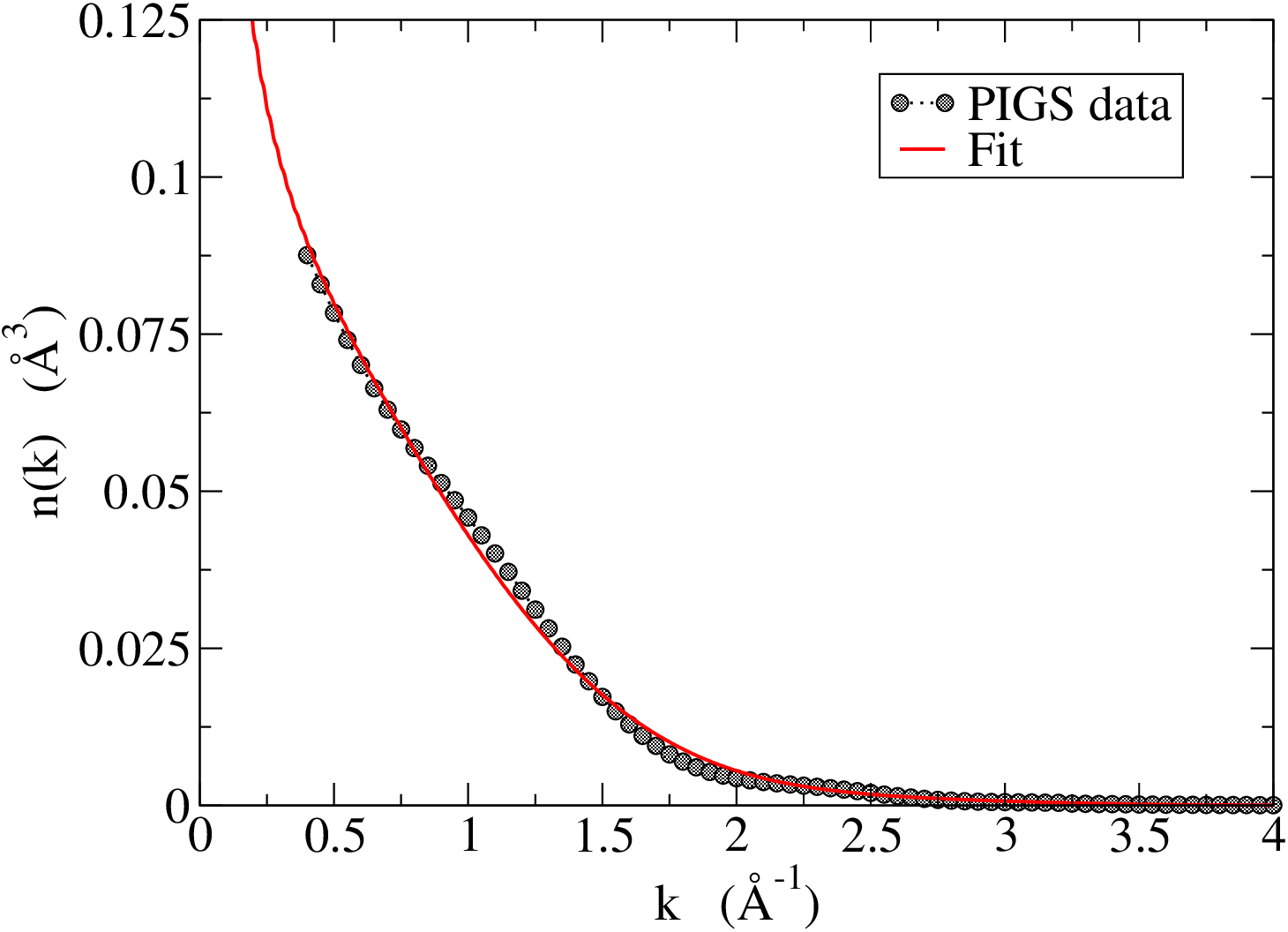}
\caption{(Color online) The momentum distribution $n(k)$ at equilibrium density
$\rho = 0.02186$ \AA$^{-3}$: the black circles represents the
numerical result obtained from the PIGS simulation, the red line
represents the FT of the fit for $\rho_1(r)$ obtained according to
Eq. (\ref{Eq_OBDM_fit}).}\label{Fig_nk_eq}
\end{figure}

In Fig. \ref{Fig_nk_eq}, we show results of $n(k)$ obtained performing a numerical Fourier transform of
$\rho_1(r)$ and we compare it with the Fourier transform of  Eq.
\ref{Eq_OBDM_fit},
\begin{equation}
n(k) = n_0 \delta(k) + n_0 f(k) + A n^*(k) \ .
\label{fullnk}
\end{equation}
The PIGS data are plotted from $k_{min} = 2\pi/L \simeq
0.4 \,$ \AA$^{-1}$ , $L$ being the length  of the simulation box, and
are not able to reproduce the $1/k$ behavior of $n(k)$ at low $k$ because
of finite size  effects; for $k > k_{min}$ the effect of $f(k)$ vanishes
and $n(k)=n^*(k)$.
We notice that the disagreement between the two curves is larger in the region
between $k \simeq 1 \,$ \AA$^{-1}$ and $k \simeq 2.5 \,$
\AA$^{-1}$. In this range, indeed, $n(k)$ obtained with the
PIGS method presents a change of curvature, not seen in $n(k)$
obtained from the fit.

This discrepancy can be explained considering the coupling between the
condensate and the states out of it. The term $f(k)$, defined in Eq.
(\ref{Eq_fk}), is obtained considering only pure density excitations in the
system and, therefore, is valid only in the limit of small
momenta.\cite{GlydeBook} At higher $k$, one should consider even the
contributions due to the coupling of the condensate to the excited states
out of the phonon region. However, little is known about these
contributions and it is difficult to include them in a more complete form
for $f(k)$ in order to give a more reliable model for the momentum
distribution.

\begin{figure}
\includegraphics[width=\linewidth]{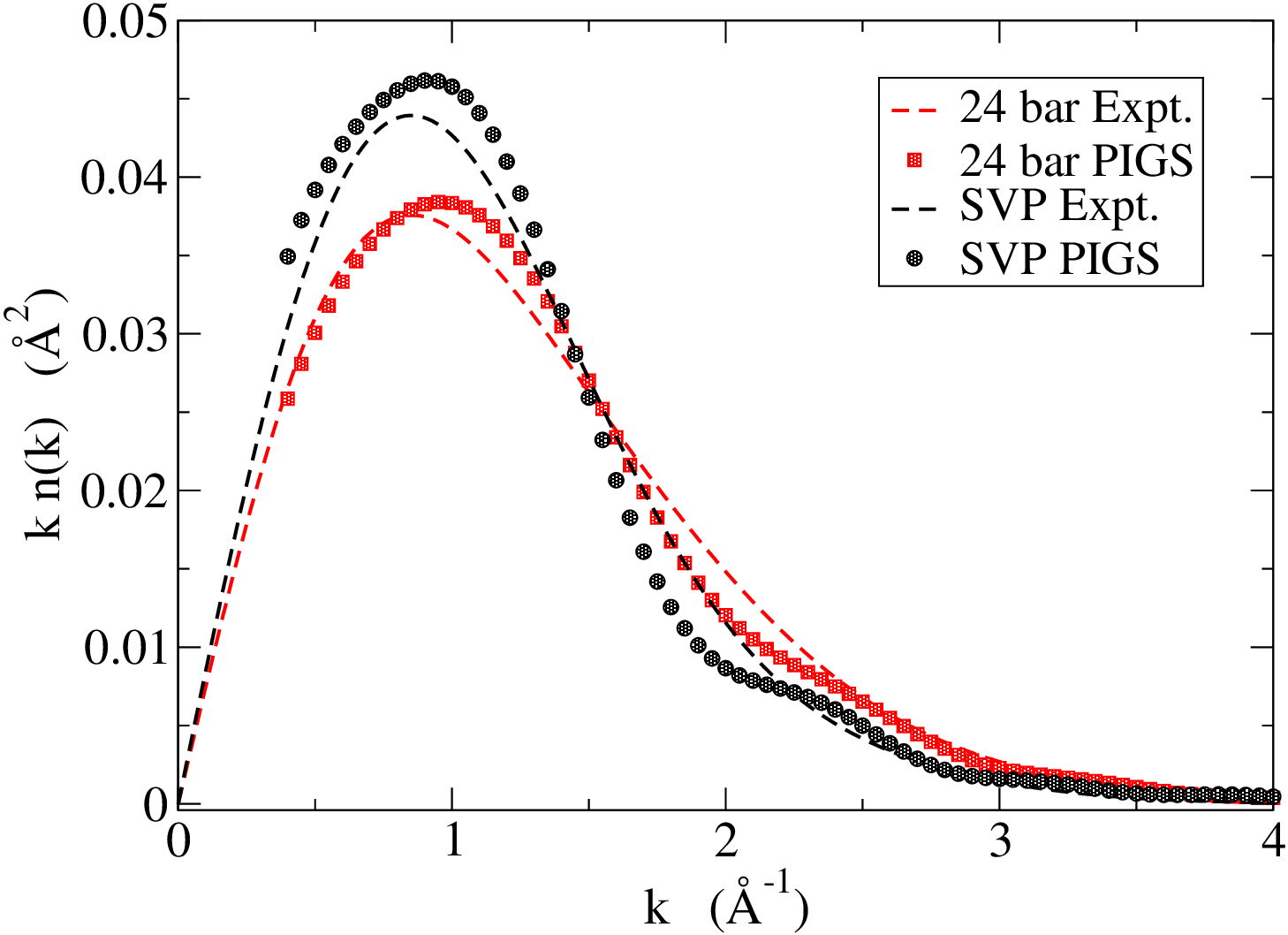}
\caption{(Color online) The momentum distribution, plotted as $k n(k)$,
in liquid $^4$He at two different pressure: black circles and red squares
are the PIGS results for $n(k)$, respectively, at saturated vapor pressure
($\rho = 0.02186 \,$ \AA$^{-3}$) and at a pressure close to the freezing,
$p \simeq 24 \, {\rm bar}$ ($\rho = 0.02539 \,$ \AA$^{-3}$). The black
and the red dashed lines represent the experimental results for the momentum
distribution above the condensate $n^*(k)$ at the same
pressures.\cite{Glyde11}}\label{Fig_nk_press}
\end{figure}

Our results for $n(k)$ are compared with recent experimental
measurements at $T = 0.06 \, {\rm K}$ of the momentum distribution
$n^*(k)$ for states above the condensate in Fig.
\ref{Fig_nk_press}. Even in this case, we can notice a good
agreement between the two curves, except for the intermediate
range of $k$, where our results include contributions arising from
the coupling between the condensate and excited states. From the
comparison between the two curves, we can deduce that this
coupling contributes in depleting the states at higher $k$. In the
same figure, in addition to $n(k)$ at the saturated vapor
pressure, we also show $n(k)$ for a higher pressure, close to the
freezing transition. We can see that the effect of the pressure in
the momentum distribution is to decrease the occupancy of the
low-momenta states and to make smoother the shoulder at $k \simeq
2 \,$ \AA$^{-1}$.

\begin{table}
 \newcolumntype{C}{>{\centering\arraybackslash}X}
 \begin{tabularx}{\linewidth}{CCCC}
 \hline
 \hline
  $\rho[$ \AA$^{-3}]$ & $p$[bar] & $n_0$ & $K/N$[K] \\
  \hline
  0.01964 & -6.23 & 0.1157(19) & 12.01(3) \\
  0.02186 & -0.04 & 0.0801(22) & 14.37(3) \\
  0.02264 & 3.29 & 0.0635(16) & 15.35(3) \\
  0.02341 & 7.36 & 0.0514(16) & 16.28(3) \\
  0.02401 & 11.07 & 0.0436(11) & 17.02(4) \\
  0.02479 & 16.71 & 0.0350(7) & 18.08(4) \\
  0.02539 & 21.76 & 0.0333(8) & 18.82(5) \\
  0.02623 & 29.98 & 0.0278(8) & 19.99(4) \\
  0.02701 & 38.95 & 0.0208(6) & 21.08(5) \\
  0.02785 & 50.23 & 0.0155(6) & 22.24(4) \\
  0.02869 & 63.37 & 0.0115(4) & 23.65(4) \\
  0.02940 & 76.28 & 0.0093(4) & 24.83(5) \\
  0.02994 & 87.06 & 0.0083(4) & 25.61(6) \\
  \hline
  \hline
\end{tabularx}
\caption{Condensate fraction $n_0$ and kinetic energy per particle $K/N$ as
a function of the liquid density $\rho$. Numbers in parenthesis are
statistical errors. The estimation of the pressure is obtained from the
equation of state of liquid $^4$He given in
Ref.~\onlinecite{Vranjes05}}\label{Tab_n0}
\end{table}

Finally, we report our results for the condensate fraction $n_0$
over a wide range of densities, including also densities in the
negative pressure region and in the regime of the overpressurized
metastable fluid. In this range of high densities, we have been
able to frustrate the formation of the crystal by starting the
simulation from an equilibrated disordered configuration. The
metastability of this phase is checked by monitoring how the total
energy per particle $E/N$ changes with the number of Monte Carlo
steps. As the simulation goes on, we notice that $E/N$
reaches a plateau for a value above the corresponding value of $E/N$
computed in a perfect crystal at the same density. For instance, at 
density $\rho = 0.02940 \,$ \AA$^{-3}$ we get in our simulation $E/N =
(-5.48 \pm 0.03) \, {\rm K}$. If we perform a PIGS simulation at the same density and
with the same choice for the initial trial wave function (in both cases, we choose in Eq. \ref{Eq_PIGSwf} $\Psi_T = 1$) but
starting the computation from a hcp crystalline configuration, we
get $E/N = (-5.95 \pm 0.02) \, {\rm K}$. The disagreement of the two results for $E/N$ indicates that, in PIGS
simulations, initial conditions for the atomic configuration
influence the evolution of the system: in particular, a sensible
choice of the initial conditions speed up the convergence of the
system to the real equilibrium state. In the simulation of $^4$He
at high densities, if we use a disordered configuration as the
initial one, the system evolves towards the equilibrium crystalline phase, but, since
crystallization is a very slow process in PIGS simulations, we see
that the overpressurized liquid phase is metastable for a number of Monte
Carlo steps sufficiently large to give good statistics for the
ground state averages of the physical observables. If the density is
increased even more ($p>90$ bars), one starts to observe the formation of
crystallites and the stabilization of the liquid becomes more difficult.

Another evidence of the metastability of the liquid configuration
in our simulations can be given computing the static structure
factor $S(k)$. In all the calculations performed, we notice the
absence of Bragg peaks in $S(k)$, which indicates clearly that the
system does not present crystalline order.

Our results for $n_0$ at different $p$ are contained in Table
\ref{Tab_n0}, together with our estimates for the kinetic energy
$K/N$. It is interesting to notice that the condensate fraction of
the overpressurized liquid is finite also for densities above the
melting ($\rho \ge 0.02862 \,$ \AA$^{-3}$). This evidence supports
our hypothesis that the system has reached a metastable
non-crystalline phase, since recent PIGS simulations show that, in
commensurate hcp $^4$He crystals, the one-body density matrix
decays exponentially to zero at large distances and therefore BEC
is not present\cite{Galli08,Rota11}. In particular, we obtain that
in the overpressurized fluid at the melting density the condensate
fraction is $n_0 \simeq 1.2\%$. This result, even though cannot
provide any deeper understanding concerning the quest of
supersolidity in $^4$He,\cite{Galli08} can be thought as an upper
limit for the condensate fraction in solid $^4$He at melting. It
is also interesting to notice that, even at the freezing pressure,
the condensate fraction is already quite small, $n_0 = 2.9\%$.

\begin{figure}
\includegraphics[width=\linewidth]{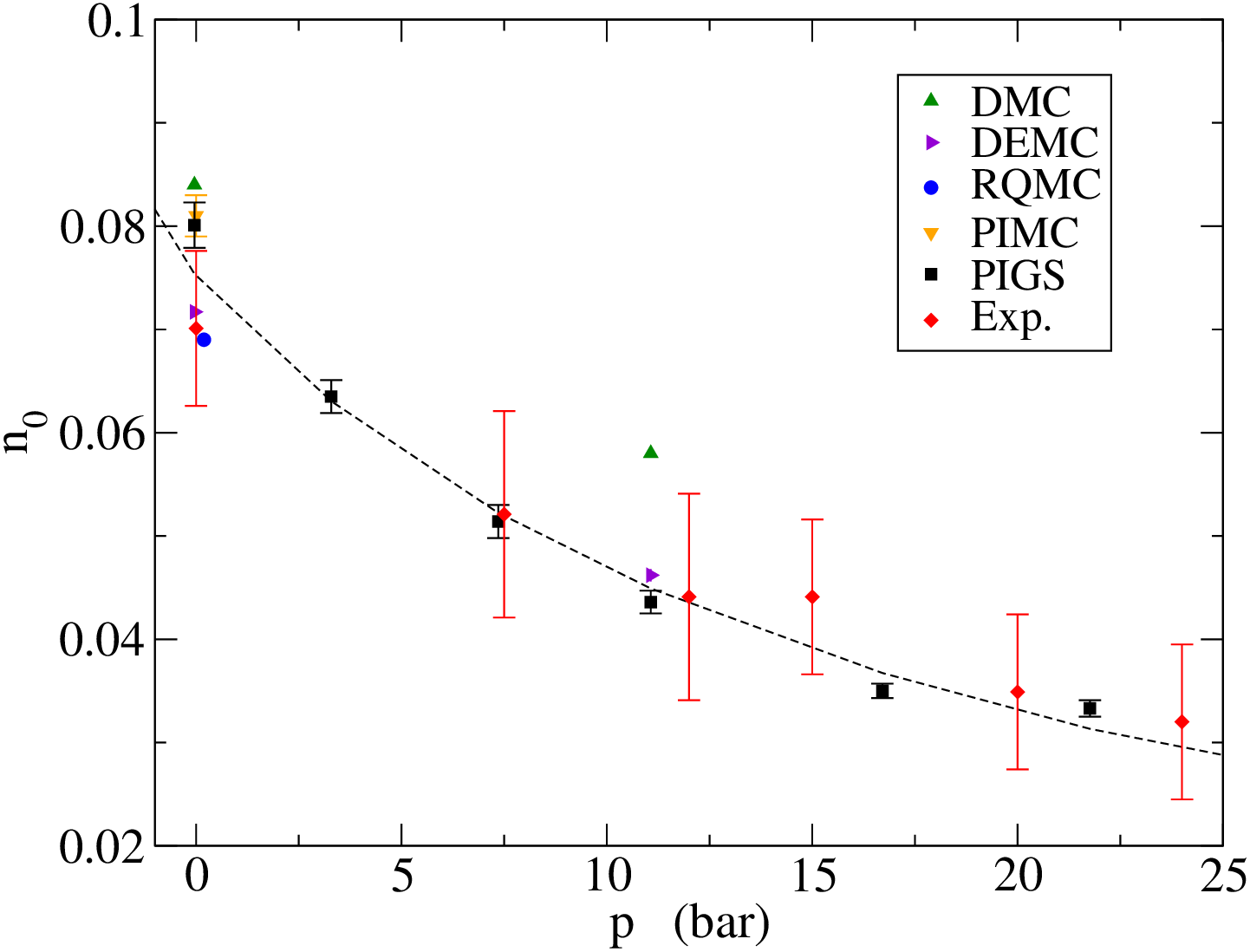}
\caption{(Color online) The condensate fraction $n_0$ in liquid
$^4$He at zero temperature as a function of pressure $p$ in the
region of stability of liquid phase. Our PIGS results (black
squares) are compared with the experimental ones (red diamonds)
\cite{Glyde11} and previous theoretical calculations, obtained
with Diffusion Monte Carlo (green up triangles),\cite{Boronat94}
Diffusion Euler Monte Carlo (violet left
triangles),\cite{Moroni97} Reptation Quantum Monte Carlo (blue
circle) \cite{Moroni04} and PIMC at $T = 1 \, {\rm
K}$.\cite{BoninsegniWorm} The dashed line represents the curve
obtained fitting our results with the equation $n_0 = A +
B/(p-p_0)$.}\label{Fig_n0_press}
\end{figure}

In Fig. \ref{Fig_n0_press}, we plot our results for $n_0$ as a
function of $p$ on the range of pressures where the liquid phase
is stable. Our results follow well an inverse proportionality law
$n_0(p) = A + B/(p-p_0)$, with $p$ and $p_0$ measured in bar: the
best fit we got has parameters $A = -0.0068 \pm 0.0012$, $B = 1.56
\pm 0.10$, $p_0 = -19.0 \pm 0.9 \, {\rm bar}$. In Fig.
\ref{Fig_n0_press}, we also compare our estimates for $n_0$ with
the experimental ones\cite{Glyde11} and with the ones obtained in
previous numerical
simulations.\cite{Boronat94,Moroni97,Moroni04,BoninsegniWorm} It
is easy to notice that our results provide an excellent
description of the experimental dependence of the condensate
fraction as a function of pressure in all the range of stability
of the liquid phase of $^4$He, improving previous calculations
which focus especially on the equilibrium density and do not
explore in detail the physically interesting pressure range where
the experimental data can be measured. Notice that the experimental value
of $n_0$ at zero pressure reported in the more recent experiment~\cite{Glyde11} 
is slightly smaller ($7.01 \pm 0.75$\%), but still statistically compatible 
within the error bars, than the previous one by the same team.~\cite{Glyde00}

\section{Conclusions}\label{Sec_Conclusions}
We have computed the one-body density matrix of liquid $^4$He at zero
temperature and different densities by means of the PIGS Monte Carlo
method. Although it is not easy to give an analytic model to fit the data
for $\rho_1(r)$, because of the difficulty of describing the coupling
between the condensate and the excited states in strongly correlated
quantum systems such as $^4$He, it is possible to extrapolate very precise
estimates of the condensate fraction and of the kinetic energy of the
system even from a simplified model for $\rho_1$. Our calculations provide
an improvement with respect to the other ground state projection techniques
used in the past, since the PIGS method allows us to remove completely the
influence of any input trial wave function. Indeed, we have performed
calculations of $\rho_1(r)$ and $n_0$ in liquid $^4$He
at zero temperature using a model for the ground state wave function which
depends uniquely on the Hamiltonian and on the symmetry properties of the
system. At the equilibrium density of liquid $^4$He, we have recovered the
value of $n_0$ obtained with the unbiased PIMC method at temperature $T
= 1 \, {\rm K}$.\cite{BoninsegniWorm} Simulating the system at several
densities, the dependence of $n_0$ with pressure $p$ obtained from the
calculation agrees nicely with the recent experimental
measurements of Ref.~\onlinecite{Glyde11}

\begin{acknowledgements}
Authors would like to thank Henry Glyde for helpful discussion and for
sending us unpublished data. This work was partially supported by DGI
(Spain) under Grant No. FIS2008-04403 and Generalitat de Catalunya under
Grant No. 2009-SGR1003.
\end{acknowledgements}

% \pagebreak
\appendix
\section{Appendix: full tables for the momentum distribution}

The aim of this appendix is to provide the full table of the momentum distribution $n(k)$ in liquid $^4$He at zero temperature and different densities, obtained from Path Integral Ground State caluclations.

\begin{table}[h]
 \newcolumntype{C}{>{\centering\arraybackslash}X}
 \begin{tabularx}{\textwidth}{CCCCCCC}
 \hline
 \hline
& & & & & & \\
$\rho $   (\AA$^{-3}$)&  0.01964 & 0.02186 & 0.02264 & 0.02341 & 0.02401 & 0.02479 \\
& & & & & & \\
\hline
& & & & & & \\
k  (\AA$^{-1}$)  & \multicolumn{6}{c}{n(k)   (\AA$^3$)}\\
& & & & & & \\
0.4 &  0.10415 &  0.08758 &  0.08041 &  0.07708 &  0.07685 &  0.07078 \\
0.5 &  0.09226 &  0.07838 &  0.07355 &  0.07082 &  0.06972 &  0.06509 \\
0.6 &  0.08157 &  0.07009 &  0.06697 &  0.06448 &  0.06292 &  0.05937 \\
0.7 &  0.07238 &  0.06298 &  0.06092 &  0.05833 &  0.05676 &  0.05390 \\
0.8 &  0.06441 &  0.05687 &  0.05532 &  0.05244 &  0.05127 &  0.04877 \\
0.9 &  0.05711 &  0.05130 &  0.04996 &  0.04682 &  0.04628 &  0.04397 \\
1.0 &  0.04996 &  0.04581 &  0.04461 &  0.04141 &  0.04154 &  0.03941 \\
1.1 &  0.04269 &  0.04011 &  0.03912 &  0.03619 &  0.03684 &  0.03499 \\
1.2 &  0.03536 &  0.03417 &  0.03351 &  0.03116 &  0.03209 &  0.03066 \\
1.3 &  0.02823 &  0.02818 &  0.02793 &  0.02637 &  0.02733 &  0.02644 \\
1.4 &  0.02165 &  0.02244 &  0.02262 &  0.02186 &  0.02272 &  0.02238 \\
1.5 &  0.01596 &  0.01727 &  0.01781 &  0.01771 &  0.01843 &  0.01858 \\
& & & & & & \\
1.6 &  0.01135 &  0.01291 &  0.01367 &  0.01401 &  0.01465 &  0.01514 \\
1.7 &  0.00792 &  0.00948 &  0.01032 &  0.01082 &  0.01149 &  0.01215 \\
1.8 &  0.00561 &  0.00699 &  0.00777 &  0.00824 &  0.00898 &  0.00965 \\
1.9 &  0.00425 &  0.00533 &  0.00594 &  0.00630 &  0.00709 &  0.00767 \\
2.0 &  0.00356 &  0.00433 &  0.00472 &  0.00495 &  0.00573 &  0.00617 \\
2.1 &  0.00325 &  0.00374 &  0.00393 &  0.00408 &  0.00476 &  0.00506 \\
2.2 &  0.00303 &  0.00334 &  0.00341 &  0.00354 &  0.00406 &  0.00424 \\
2.3 &  0.00272 &  0.00296 &  0.00300 &  0.00314 &  0.00350 &  0.00361 \\
2.4 &  0.00225 &  0.00252 &  0.00260 &  0.00276 &  0.00300 &  0.00308 \\
2.5 &  0.00170 &  0.00201 &  0.00217 &  0.00232 &  0.00251 &  0.00258 \\
& & & & & & \\
2.6 &  0.00118 &  0.00149 &  0.00171 &  0.00184 &  0.00204 &  0.00210 \\
2.7 &  0.00078 &  0.00106 &  0.00128 &  0.00138 &  0.00161 &  0.00166 \\
2.8 &  0.00055 &  0.00077 &  0.00092 &  0.00100 &  0.00124 &  0.00127 \\
2.9 &  0.00045 &  0.00061 &  0.00065 &  0.00074 &  0.00095 &  0.00097 \\
3.0 &  0.00040 &  0.00053 &  0.00048 &  0.00058 &  0.00075 &  0.00076 \\
3.1 &  0.00036 &  0.00048 &  0.00038 &  0.00049 &  0.00062 &  0.00062 \\
3.2 &  0.00029 &  0.00042 &  0.00032 &  0.00041 &  0.00052 &  0.00053 \\
3.3 &  0.00021 &  0.00027 &  0.00027 &  0.00032 &  0.00044 &  0.00046 \\
3.4 &  0.00016 &  0.00021 &  0.00022 &  0.00023 &  0.00037 &  0.00039 \\
3.5 &  0.00015 &  0.00017 &  0.00017 &  0.00015 &  0.00030 &  0.00031 \\
\hline
\hline
\end{tabularx}
\caption{The momentum distribution $n(k)$ of liquid $^4$He at zero temperature as computed by Path Integral Ground State method for densities between $\rho = 0.01964 \, {\rm \AA}^{-3}$ and  $\rho = 0.02479 \, {\rm \AA}^{-3}$}\label{Tab1}
\end{table}

\begin{table}
 \newcolumntype{C}{>{\centering\arraybackslash}X}
 \begin{tabularx}{\textwidth}{CCCCCCCC}
 \hline
 \hline
& & & & & & & \\
$\rho $   (\AA$^{-3}$)& 0.02539 & 0.02623 & 0.02701 & 0.02785 & 0.02869 & 0.02940 & 0.02994 \\
& & & & & & & \\
\hline
& & & & & & & \\
k  (\AA$^{-1}$)  & \multicolumn{7}{c}{n(k)   (\AA$^3$)}\\
& & & & & & & \\
0.4 &  0.06464 &  0.06170 &  0.05811 &  0.05400 &  0.04904 &  0.04542 & 0.04310 \\
0.5 &  0.06013 &  0.05751 &  0.05403 &  0.04993 &  0.04600 &  0.04291 & 0.04076 \\
0.6 &  0.05554 &  0.05328 &  0.04990 &  0.04589 &  0.04286 &  0.04025 & 0.03828 \\
0.7 &  0.05106 &  0.04916 &  0.04590 &  0.04210 &  0.03976 &  0.03753 & 0.03578 \\
0.8 &  0.04673 &  0.04520 &  0.04211 &  0.03862 &  0.03677 &  0.03483 & 0.03330 \\
0.9 &  0.04252 &  0.04134 &  0.03849 &  0.03541 &  0.03389 &  0.03217 & 0.03086 \\
1.0 &  0.03834 &  0.03752 &  0.03499 &  0.03239 &  0.03107 &  0.02954 & 0.02843 \\
1.1 &  0.03416 &  0.03367 &  0.03153 &  0.02942 &  0.02830 &  0.02694 & 0.02602 \\
1.2 &  0.02995 &  0.02980 &  0.02807 &  0.02644 &  0.02552 &  0.02435 & 0.02359 \\
1.3 &  0.02579 &  0.02594 &  0.02463 &  0.02343 &  0.02276 &  0.02178 & 0.02118 \\
1.4 &  0.02177 &  0.02219 &  0.02127 &  0.02043 &  0.02003 &  0.01927 & 0.01879 \\
1.5 &  0.01801 &  0.01864 &  0.01806 &  0.01751 &  0.01740 &  0.01684 & 0.01649 \\
& & & & & & & \\
1.6 &  0.01462 &  0.01539 &  0.01510 &  0.01479 &  0.01491 &  0.01456 & 0.01430 \\
1.7 &  0.01171 &  0.01253 &  0.01247 &  0.01234 &  0.01264 &  0.01245 & 0.01229 \\
1.8 &  0.00931 &  0.01010 &  0.01021 &  0.01023 &  0.01063 &  0.01056 & 0.01048 \\
1.9 &  0.00743 &  0.00814 &  0.00835 &  0.00848 &  0.00890 &  0.00892 & 0.00891 \\
2.0 &  0.00602 &  0.00662 &  0.00687 &  0.00707 &  0.00746 &  0.00752 & 0.00756 \\
2.1 &  0.00500 &  0.00549 &  0.00571 &  0.00595 &  0.00629 &  0.00636 & 0.00644 \\
2.2 &  0.00425 &  0.00466 &  0.00481 &  0.00506 &  0.00533 &  0.00540 & 0.00550 \\
2.3 &  0.00365 &  0.00402 &  0.00410 &  0.00433 &  0.00455 &  0.00461 & 0.00473 \\
2.4 &  0.00312 &  0.00348 &  0.00350 &  0.00370 &  0.00388 &  0.00395 & 0.00407 \\
2.5 &  0.00261 &  0.00296 &  0.00297 &  0.00314 &  0.00330 &  0.00337 & 0.00349 \\
& & & & & & & \\
2.6 &  0.00211 &  0.00246 &  0.00248 &  0.00262 &  0.00278 &  0.00286 & 0.00297 \\
2.7 &  0.00165 &  0.00198 &  0.00203 &  0.00215 &  0.00230 &  0.00240 & 0.00250 \\
2.8 &  0.00126 &  0.00155 &  0.00162 &  0.00173 &  0.00188 &  0.00200 & 0.00208 \\
2.9 &  0.00096 &  0.00121 &  0.00129 &  0.00139 &  0.00152 &  0.00164 & 0.00171 \\
3.0 &  0.00076 &  0.00097 &  0.00102 &  0.00111 &  0.00122 &  0.00134 & 0.00140 \\
3.1 &  0.00064 &  0.00081 &  0.00081 &  0.00090 &  0.00100 &  0.00111 & 0.00114 \\
3.2 &  0.00055 &  0.00070 &  0.00066 &  0.00074 &  0.00082 &  0.00092 & 0.00094 \\
3.3 &  0.00047 &  0.00061 &  0.00054 &  0.00062 &  0.00069 &  0.00077 & 0.00078 \\
3.4 &  0.00039 &  0.00052 &  0.00044 &  0.00053 &  0.00058 &  0.00065 & 0.00066 \\
3.5 &  0.00031 &  0.00043 &  0.00035 &  0.00044 &  0.00049 &  0.00055 & 0.00055 \\
\hline
\hline
\end{tabularx}
\caption{The momentum distribution $n(k)$ of liquid $^4$He at zero temperature as computed by Path Integral Ground State method for densities between $\rho = 0.02539 \, {\rm \AA}^{-3}$ and  $\rho = 0.02994 \, {\rm \AA}^{-3}$}\label{Tab2}
\end{table}

\end{document}